\documentclass[%
 reprint,
 footinbib,
 amsmath,amssymb,
 aps,
prl,
]{revtex4-1}
\usepackage{appendix}
\usepackage{graphicx}% Include figure files
\usepackage{dcolumn}% Align table columns on decimal point
\usepackage{bm}% bold math
\usepackage{tabularx}
\usepackage{pdfpages}
\makeatletter
\AtBeginDocument{\let\LS@rot\@undefined}
\makeatother
\usepackage{setspace}
\begin{document}
\renewcommand{\arraystretch}{2.0}
\renewcommand{\baselinestretch}{1}

%\preprint{APS/123-QED}

\title{Statistical Mechanics of Low Angle Grain Boundaries in Two Dimensions}% Force line breaks with \\

\author{Grace H. Zhang}
%\email{ghzhang@g.harvard.edu}
\affiliation{Department of Physics, Harvard University, Cambridge, MA 02138, USA.}
\author{David R. Nelson}
\affiliation{Department of Physics, Harvard University, Cambridge, MA 02138, USA.}%

\date{\today}
% \pagenumbering{gobble}

\begin{abstract}
We explore order in low angle grain boundaries (LAGBs) embedded in a two-dimensional crystal at thermal equilibrium. Symmetric LAGBs subject to a Peierls potential undergo, with increasing temperatures, a thermal depinning transition, above which the LAGB exhibits transverse fluctuations that grow logarithmically with inter-dislocation distance. Longitudinal fluctuations lead to a series of melting transitions marked by the sequential disappearance of diverging algebraic Bragg peaks with universal critical exponents. Aspects of our theory are checked by a mapping onto random matrix theory. 
\end{abstract}

\pacs{Valid PACS appear here}% PACS, the Physics and Astronomy
% \tableofcontents             
% \clearpage
\maketitle

Grain boundaries, interfaces dividing crystal grains with distinct orientations, significantly impact the properties of all polycrystalline materials~\cite{gottstein2009grain, balluffi2006interfaces, cantwell2014grain}. Their dynamics directly affect grain growth stagnation~\cite{holm2010grain}, grain boundary mobility~\cite{olmsted2007grain}, superplasticity~\cite{lu2000superplastic}, and shear strength~\cite{broughton1998grain,cahn2006coupling}, altering the microstructure evolution of a wide class of materials~\cite{cantwell2014grain}, including high-$T_c$ superconductors~\cite{hilgenkamp2002grain} and two-dimensional (2d) materials~\cite{limbu2017grain,zhang2013mechanical,yazyev2010electronic}. 
A complete understanding of grain boundary dynamics is crucial for informing materials synthesis methods and industrial processes~\cite{cantwell2014grain,babcock1995nature, yu2011control,schweizer2018situ}.

While previous works have focused on the roughening~\cite{rottman1986roughening,chui2009grain,hsieh1989observations,lee2018roughening,liao2018grain}, defaceting~\cite{daruka2004atomistic,olmsted2007grain}, 
premelting~\cite{kikuchi1980grain,alsayed2005premelting}, and structural phase transitions~\cite{olmsted2011dislocation,frolov2013structural} in three-dimensional crystals,
recent advances in colloid science~\cite{li2016assembly} and 2d electronic devices~\cite{novoselov20162d,zhang2017two,van2013grains} have triggered studies of grain boundaries in 2d crystals, with electronic and thermal properties especially sensitive to lattice imperfections~\cite{mellenthin2008phase,moretti2004depinning,leoni2007grain,lipowsky2005direct,bausch2003grain,skinner2010grain,huang2011grains, kim2011grain,gibb2013atomic}. 

\begin{figure}[tb]
    \centering
    \includegraphics[width=0.85\columnwidth]{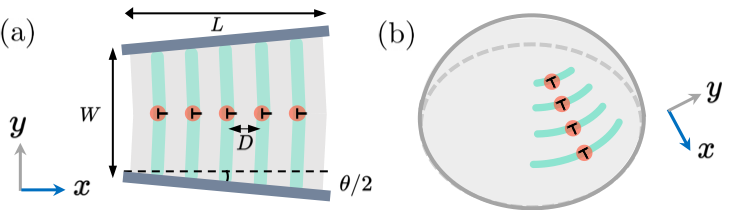}
    \caption{Schematic of a single LAGB consisting of point dislocations (orange) with Burgers vectors aligned along their glide planes (turquoise), embedded in a flat 2d crystal with wedge angle $\theta$ (a) and in a curved 2d crystal (b). }
    \label{fig:LAGB_schm}
\end{figure}

\begin{figure}[htb]
    \centering
    \includegraphics[width=0.9\columnwidth]{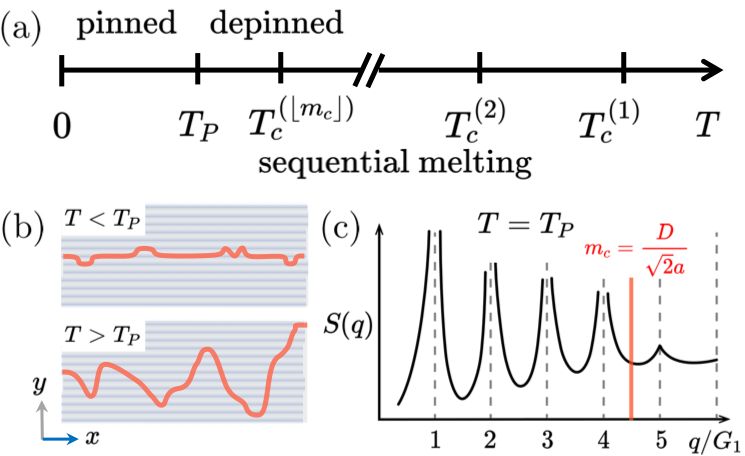}
    \caption{(a): Phase diagram of a LAGB embedded in a 2d crystal. (b): Schematic of an LAGB below and above the depinning transition, with transverse grain boundary correlations given by Eq.~(\ref{eq:corr_pin}) and (\ref{eq:corr_depin}). Gray lines illustrates the transverse Peierls potential. (c): Schematic of the structure factor ($q \equiv q_x)$) at the depinning transition with power law divergences in Bragg peaks according to Eqs.~(\ref{eq:Sq_final}) at $\{G_m \}$ provided the Bragg peak index  $m < m_c$ (Eq.~(\ref{eq:mc})). }
    \label{fig:LAGB_pd}
\end{figure}

In two dimensions, grain boundaries have been proposed as a mechanism for two dimensional melting~\cite{chui1982grain,chui1983grain}. The free energy of low angle grain boundaries (LAGBs) changes sign at the Kosterlitz-Thouless melting temperature~\cite{fisher1979defects}. However, much less is known about their statistical mechanics at lower temperatures, particularly in the presence of a periodic Peierls pinning potential. In this work, we study the statistical mechanics of LAGBs embedded in a host 2d crystal. 
Although LAGBs may not be an equilibrium feature of crystals with rectangular boundary conditions, they appear in the ground state of flat crystals with e.g. trapezoidal boundary conditions. See Fig.~\ref{fig:LAGB_schm}a, where a crystal with length $L$ and width $W$ is trapped between slanted walls, as could be studied in both experiments and simulations. When $W \sim L \gg a$, it is straightforward to show that a grain boundary is preferred over a strained, defect-free crystal (see Supplemental Material~\footnote{\label{supp} See Supplemental Material at \texttt{link}, which includes Refs.~\cite{kosterlitz2017nobel,wigner1958distribution,livan2018introduction,zhang2020pileups}.}). 
We model the LAGB as a one-dimensional array of dislocations, with identical Burgers vectors directed perpendicular to the boundary, embedded in a 2d continuous elastic medium. ``Low angle'' means dislocation spacings large enough so that these defects are well-defined, with glide planes approximately perpendicular to the interface itself. We also assume a periodic Peierls potential transverse to the boundary itself~\cite{hirth1983theory} (Fig.~\ref{fig:LAGB_schm}). Mapping onto a model of quantum Brownian motion in imaginary time~\cite{fisher1985quantum} allows a renormalization group treatment of the depinning transition. By analytically calculating dislocation correlations in various regimes, and numerically testing our theory using a mapping onto random matrix theory, we uncover the phase diagram in Fig.~\ref{fig:LAGB_pd}a. The sharp one-dimensional phase transitions displayed in Fig.~\ref{fig:LAGB_pd}a are only possible because of the long range interactions between dislocations in the LAGB. 
Note that the transitions in Eqs.~(\ref{eq:TR}) and (\ref{eq:Tcm}) below can also be expressed in terms of a dimensionless temperature $\Gamma \equiv \frac{k_B T}{Yb^2}$, so the transitions can also be realized by tuning the interaction strength instead of $k_B T$. 

At low temperatures, the LAGB is in a pinned phase, with dislocations  localized along their glide directions by Peierls potential. Upon increasing the temperature, we find a depinning transition at $T = T_P$, above which the Peierls potential is overcome by thermal fluctuations and the transverse LAGB fluctuations grow logarithmically with distance along the boundary. As the temperature continues to increase, the quasi-long range ordered depinned LAGB melts via a series of phase transitions at $\{T_c^{(m)} \}$, marked by the sequential disappearance of power law divergences at ever-smaller Bragg peaks $\{ G_m \}$. These transitions proceed until only the final peak at $G_1$ remains, after which the LAGB melts with the host crystal, if the host crystal has not already melted by some other mechanism (see, e.g., Ref.~\cite{li2020attraction} and references therein). 

Equilibrium configurations of symmetric grain boundaries in $d=2$, without shear load, can be modelled by point-like edge dislocations with glide planes perpendicular to the boundary (Fig.~\ref{fig:LAGB_schm}a). Note that LAGBs can also arise on curved 2d crystals~\cite{azadi2014emergent,azadi2016neutral,bowick2000interacting}, where grain boundary scars form to relieve the excess strain introduced by Gaussian curvature and associated topological defects such as disclinations~\cite{bausch2003grain,bowick2000interacting} (Fig.~\ref{fig:LAGB_schm}b).

The energy of a LAGB, consisting of $N$ point edge dislocations, embedded in a 2d crystal is~\cite{hirth1983theory},
\begin{widetext} \normalsize 
\begin{equation}
\begin{aligned}
H [  \{ x_n ,y_n \} ] &=
 - \frac{Yb^2}{8 \pi} \sum_{n \neq m} \Bigg[\frac{1}{2}   \ln \left |(x_n - x_m)^2 + (y_n - y_m)^2\right|  - \frac{ (y_n - y_m)^2}{ (x_n - x_m)^2 + (y_n - y_m)^2} \Bigg] 
 -V_{\mathrm{Peierls}} \sum_n \cos \left(\frac{2 \pi y_n}{a}\right), \label{eq:H_GB}
 \end{aligned}
\end{equation}
\end{widetext} \normalsize  
% \small  
% \begin{eqnarray} 
% H [ n ( x ,y) ] =
%  - \frac{Yb^2}{8 \pi} \sum_{n \neq m} \Bigg[\frac{1}{2}   \ln \left |[(n-m)D + (u^{(x)}_n - u^{(x)}_m)]^2 + (y-y')^2 \right| - \frac{ (u^{(y)}_n - u^{(y)}_m)^2}{[(n-m)D + (u^{(x)}_n - u^{(x)}_m)]^2 + (u^{(y)}_n - u^{(y)}_m)^2} \Bigg] 
%  -V_{\mathrm{Peierls}} \sum_n \cos \left(\frac{2 \pi u^{(y)}_n}{a}\right), \label{eq:H_GB}
% \end{eqnarray} \normalsize  
where the LAGB is oriented along $\hat x$, $b$ is the magnitude of the Burgers vector along $\hat y$ (assumed equal to the lattice constant $a$ for simplicity), and $Y = \frac{ 4 \mu(\mu + \lambda)}{2 \mu + \lambda}$ is the 2d Young's modulus, where $\mu$ and $\lambda$ are the Lam\'e coefficients. The position of the $n$-th dislocation is given by $ (x_n, y_n) = (nD + u^{(x)}_n, u^{(y)}_n)$, where $u_n^{(x)}$ and $u_n^{(y)}$ are the longitudinal and transverse displacements of these defects due to thermal fluctuations. Their equilibrium positions are aligned at $y =0$, to balance the Peach-Kohler force due to interactions with other dislocations, and evenly spaced along $\hat x$ with an average spacing $D$~\cite{hirth1983theory,moretti2004depinning,leoni2007grain}. We assume that both glide and climb displacements are in thermal equilibrium, as could be achieved by having a 2d host crystal coexisting with a 3d vapor phase, which effectively supplies a reservoir of vacancies and interstitials. 

Provided that $T$ is not too close to the melting temperature $T_m$ of the host crystal,  we expect that typical displacement differences are much smaller than the dislocation separations: $|u^{}_n -u^{}_m|  \ll |n-m| D $.   
Upon expanding to quadratic order in the transverse glide displacements $\{u_n^{(y)}\}$, we obtain~\cite{moretti2004depinning,chui1983grain,leoni2007grain}
\begin{eqnarray}
H&=&\frac{Y b^{2}}{8 D^{2}} \int \frac{d q_{x}}{2 \pi}\left|q_{x}\right| \left| u^{(y)}(q_{x})\right|^{2}-V_{\text {Peierls}} \sum_{n} \cos \left(\frac{2 \pi u_{n}^{(y)}}{a}\right) \notag \\
&&-\frac{Y b^{2}}{8 \pi} \sum_{n \neq m} \ln \left|D(n-m)+(u_{n}^{(x)}-u_{m}^{(x)})\right|. 
\end{eqnarray}The first term, representing long range interactions, has been written in Fourier space. 
% Upon taking derivatives of Eq.~(\ref{eq:H_GB}) with respect to density $n(\vec r) = \sum_n \delta(\vec r - \vec r_n)$ in the continuum limit and the spatial coordinate $y$, , we obtain the force balance condition $0 = f_{y}\left(x, y\right) $ as~\cite{hirth1983theory}:\footnotesize  
% \begin{eqnarray} 
% 0 &=& -\frac{Yb^2}{8 \pi} \sum_n \frac{\left(y-y'\right)\left[\left(y-y'\right)^{2}-\left(x-x'\right)^{2}\right]}{\left[\left(y-y'\right)^{2}+\left(x-x'\right)^{2}\right]^{2}}. \label{eq:fb_gb}
% \end{eqnarray} \normalsize  
% which is satisfied if the dislocations are condensed onto a single value of $y$:
% \begin{eqnarray}
% n(x, y) = n(x) \delta(y-y_0),
% \end{eqnarray}
% where $n(x) = \sum_j \delta(x-x_j)$, given a set of dislocation coordinates $\{ x_j \}$, and $y_0$ is a constant fixed the locations of triple junctions at the ends of the LAGB~\cite{hirth1983theory}. Hence forth, we'll set $y_0 = 0$ without loss of generality. 

\textit{Thermal depinning.}---We first examine the behavior of glide fluctuations transverse to the boundary. Upon integrating out the longitudinal $\{u_n^{(x)}\}$ degrees of freedom, the fluctuation energy for the remaining transverse modes becomes \footnotesize  
\begin{eqnarray}
H_\text{t} &=&\frac{Y b^{2}}{8 D^{2}} \int \frac{d q_{x}}{2 \pi}\left|q_{x}\right|\left|u^{(y)}\left(q_{x}\right)\right|^{2} -V_{\mathrm{Peierls}} \sum_n \cos \left(\frac{2 \pi u_n^{(y)}}{a}\right).
\label{eq:LAGB_H_QBM} 
\end{eqnarray} \normalsize  
A similar Hamiltonian was proposed to model dislocations in a zipper-like interface~\cite{kolomeisky1996phase}. In that case, however, the Peierls potential acts \textit{along} the boundary, rather than perpendicular to it. The reduced Hamiltonian of Eq.~(\ref{eq:LAGB_H_QBM}) maps exactly onto the action of a quantum Brownian particle in a periodic potential in imaginary time~\cite{fisher1985quantum,leggett1987dynamics,zhang2017phase}. The repulsive logarithmic interaction between dislocations maps onto a friction force that resists the quantum particle tunneling between minima in the periodic potential. The renormalization group recursion relation for the scale-dependent Peierls potential $V_\text{Peierls}(l)$ reads~\cite{fisher1985quantum,leggett1987dynamics}, 
\begin{eqnarray}
\frac{d V_\text{Peierls} (l) }{d l} &=& \left( 1 - \frac{1}{\gamma} \right) V_\text{Peierls} (l),
\end{eqnarray}
and predicts a delocalization transition for the LAGB. Here, $\gamma = \frac{1}{k_B T}\frac{Y b^2 a^2}{8 \pi D^2}$, where $a=b$ is the lattice constant of the host crystal, and we obtain the depinning temperature $T_P$ as
\begin{eqnarray}\label{eq:TR}
k_B T_P = \theta^2\frac{Yb^2}{8\pi},
\end{eqnarray}
where $\theta = a/D$ is the usual misorientation angle of a symmetric LAGB~\cite{hirth1983theory}. 

As illustrated in Fig.~\ref{fig:LAGB_pd}b, below the depinning transition $T<T_P$, the LAGB dislocations are locked close to a particular minimum of the Peierls potential even in the presence of thermal fluctuations, and the spatial correlation function of the displacements is constant for large separation distances,
\begin{eqnarray} \label{eq:corr_pin}
\lim_{x \rightarrow \infty}\left\langle\left|u^{(y)}(x)-u^{(y)}(0)\right|^{2}\right\rangle=\left(\frac{a}{2 \pi}\right)^{2} \frac{k_{B} T}{V_{\text {Peierls }}}. 
\end{eqnarray}
% The corresponding structure factor $S(q_x, q_y) = \langle | \rho(q_x, q_y) |^2 \rangle/N $ of a LAGB with $N$ dislocations exhibits delta function Bragg peaks at the reciprocal lattice vectors $q_x = G_m = m \frac{2 \pi}{D}$, each diminished by a Debye-Waller factor, \small  
% \begin{eqnarray} \label{eq:Sq_pin}
% S(q_x, q_y)  = \frac{2 \pi}{D}\sum_m\delta(q_x - G_m) e^{- m^2 \frac{a^2}{D^2} \frac{k_B T}{V_\text{Peierls}} }. 
% \end{eqnarray} \normalsize  
For $ T > T_P $, the Peierls potential can be neglected at large distances, and the spatial correlation  function of the depinned dislocation displacements grows logarithmically with the separation distances $x$ along the LAGB:
\begin{eqnarray} \label{eq:corr_depin}
\lim _{x \rightarrow \infty}\left\langle\left|u^{(y)}(x)-u^{(y)}(0)\right|^{2}\right\rangle \approx\frac{8}{\pi} \frac{k_{B} T D^{2}}{Y b^{2}} \ln (x). 
\end{eqnarray}
Although this behavior is reminiscent of roughened 2d interfaces~\cite{kardar2007statistical,chui1976phase,ohta1979xy}, typical roughened 1d interfaces with \textit{short} range interactions in fact have fluctuations that grow as $\sqrt{x}$. 

When $T \rightarrow T_P^+$, we obtain the following universal scaling relation using Eqs.~(\ref{eq:TR}) and (\ref{eq:corr_depin}),
\begin{eqnarray}
\frac{\lim _{x \rightarrow \infty}\left\langle\left|u^{(y)}(x)-u^{(y)}(0)\right|^{2}\right\rangle}{\ln(x)} = \frac{a^2 }{\pi^2}. 
\end{eqnarray}

\textit{Melting.}---
To study the melting of longitudinal LAGB order, we now focus on the climb degrees of freedom and examine the 1d structure factor as a function of momenta $q_x$ near the reciprocal lattice vectors $\{G_m = \frac{2 \pi m}{D} \}$. Both above and below the depinning transition $T= T_P$, we can integrate out the transverse degrees of freedom $\{u_n^{(y)}\}$ from the dislocation partition function associated with Eq.~(\ref{eq:H_GB}), and obtain an effective nonlinear energy for the longitudinal coordinates $\{x_n \}$ along the boundary,
\begin{eqnarray} \label{eq:H_melt}
H_\ell(\{x_n \}) &=& -\frac{Yb^2}{8 \pi} \sum_{n \neq m} \ln|x_n - x_m|. 
\end{eqnarray} 
(Neglected terms of $O((u_n^{(x)})^2 (u_n^{(y)})^2)$ do not affect the coefficient of the logarithm within perturbation theory.) On denoting $q \equiv q_x$, setting $x_n = Dn + u_n^{(x)}$, and expanding to quadratic order in longitudinal displacements, Eq.~(\ref{eq:H_melt}) in momentum space becomes ($q_x \equiv q$)~\cite{chui1983grain}
\begin{eqnarray} \label{eq:H_q}
H = \frac{Y b^{2}}{8 D^{2}} \int \frac{d q}{2 \pi}\left|q\right|\left|u^{(x)} \left(q\right)\right|^{2}. 
\end{eqnarray}
We can now compute the asymptotic forms of the structure factor $S(q) = \langle | \rho(q) |^2 \rangle /N$, where $\rho(q) = \sum_n e^{i qx_n}$, in both the $q \rightarrow 0$ and $q \rightarrow G_m$ limits. When $q \rightarrow 0$, we can construct a hydrodynamic density fluctuation field $ \delta \rho(x)$~\cite{nelson2002defects}. Upon writing the energy in Eq.~(\ref{eq:H_q}) in terms of density fluctuations with $\delta \rho(x) = \rho_0 \partial_x u^{(x)}(x) $, where $\rho_0 = \langle \rho(x) \rangle \equiv  D^{-1}$ is the average dislocation density, we  obtain from Eq.~(\ref{eq:H_q}),
\begin{eqnarray} \label{eq:Sq_0}
\lim_{q \rightarrow 0} S(q) \approx \frac{8 \pi k_B T }{ Yb^2} |\bar q|,
\end{eqnarray}
where $\bar q \equiv \frac{q}{2 \pi/D}$ is the dimensionless wavevector. 
The linear vanishing of $S(q)$ as $q \rightarrow 0$ indicates incompressibility associated with dislocations with identical Burgers vectors, similar to a Coulomb gas of like-signed charges.  

On now setting $q = G_m + k$, with $k \ll G_m$, we obtain for the structure factor $S(q)$ near the reciprocal lattice vectors $G_m \neq 0$
\begin{eqnarray} \label{eq:Sq_k}
S(q \approx G_m)  &\approx & \sum_{ s= -\infty}^\infty e^{i k sD - \frac{G_m^2}{2} \langle | u^{(x)}_s - u^{(x)}_0|^2 \rangle}. 
\end{eqnarray}
where we have used the properties of Gaussian thermal averages to evaluate $\langle \exp [ i G_m (u^{(x)}_s - u^{(x)}_0) ] \rangle$. 
On extracting $\langle u^{(x)}(q) u^{(x)}(q') \rangle$ from Eq.~(\ref{eq:H_q}) and utilizing the properties of cosine integrals~\cite{olver2010nist}, we obtain the displacement correlation $C(s) \equiv \langle |u^{(x)}_s - u^{(x)}_0|^2 \rangle$ in the limit of large $s \rightarrow \infty$ as
\begin{eqnarray} \label{eq:dispdisp}
C(s) &=&  \frac{8 D^2 k_B T  }{\pi Yb^2 }   \left( \gamma + \ln(\pi s) + O \left[ \frac{\cos(\pi s)}{s} \right] \right),
\end{eqnarray}
where $\gamma \approx 0.577 $ is the Euler–Mascheroni constant. 
Upon substituting Eq.~(\ref{eq:dispdisp}) into Eq.~(\ref{eq:Sq_k}), we obtain the singular behavior of $S(q)$ near the $m$-th reciprocal lattice vector $G_m$ as
\begin{eqnarray} \label{eq:Sq_final}
\lim_{q \rightarrow G_m} S (q)  & \sim &\frac{1}{\left |q-G_m\right|^{1-\alpha_m(T)}}. 
\end{eqnarray}
where $1-\alpha_m(T)$ is a temperature-dependent susceptibility critical exponent, 
\begin{eqnarray} \label{eq:sq_exp}
\alpha_m(T) =  m^2 \frac{16 \pi k_B T}{Yb^2}.
\end{eqnarray}
Eq.~(\ref{eq:Sq_final}) predicts that at temperatures low enough such that $\alpha_m(T) \leq 1$, the structure factor diverges as $q$ approaches the $m$-th reciprocal lattice vector $q \rightarrow G_m$, and the higher order Bragg peaks are less divergent than the more prominent ones closer to the origin. These power law Bragg peaks for LAGBs are reminiscent of the Bragg peaks below the melting temperature of 2d point particles~\cite{nelson1979dislocation,halperin1978theory}. They replace the usual Lorentzian peaks expected for conventional one-dimensional crystals without long range order~\cite{emery1978one}, and are due to long range interactions between the dislocations in the grain boundary.

As illustrated in Fig.~\ref{fig:LAGB_pd}c, the Bragg peaks at $\{ G_m \}$ (1) remain finite if $m$ is larger than a critical value $m > m_c$, or (2) diverge as a power law with exponent $1 - \alpha_m(T) $ if $m < m_c$. The critical value $m_c$ is given by 
\begin{eqnarray} \label{eq:mc}
m_c = \frac{D}{\sqrt{2} a}. 
\end{eqnarray}

As temperature increases, divergences in higher order Bragg peaks vanish sequentially at a series of transition temperatures $ \{ T_c^{(m)} \}$, where
\begin{eqnarray} \label{eq:Tcm}
k_B T^{(m)}_c = \frac{1}{m^2} \frac{Y b^2}{16 \pi}.
\end{eqnarray}
The last Bragg peak to disappear is the first-order Bragg peak at $G_1 = \frac{2 \pi}{D}$ closest to the origin in momentum space. Interestingly, the temperature at which this last Bragg peak vanishes $ T_c^{(1)}$ seems to coincide with the dislocation pair unbinding temperature of the 2d host crystal, if we neglect screening by bulk dislocation pairs. Note that the dislocation spacing $D$ drops out in Eqs.~(\ref{eq:sq_exp}) and (\ref{eq:Tcm}), because the $D$-dependence of the interaction strength in Fourier space $\sim 1/D^2$ in Eq.~(\ref{eq:H_q}) cancels against the $D$-dependence of the reciprocal lattice vectors $\{G_m \} = \{ 2 \pi m /D \}$ in Eq.~(\ref{eq:Sq_k}).

We also calculate the pair correlations embodied in the radial distribution function $g(r)$, which determines the probability of finding a second dislocation a distance $r$ away from some first existing dislocation. The quantity $g(r)$ is given by a Fourier transform of the structure factor $S(q)$~\cite{pathria2011statistical}.
% \begin{eqnarray} \label{eq:gr_Sq}
% g(r)-1 = \frac{1}{\rho_0} \int dq \left[S(q) -\delta(q) - 1\right]e^{i q r},
% \end{eqnarray}
% where $\rho_0$ is the average dislocation density. 
Since the most prominent Bragg peak at $G_1$ dominates, we obtain the long distance behavior of $g(r)$ as
\begin{eqnarray}
\lim_{r \rightarrow \infty} (g(r ) -1) \sim r^{-\alpha_1(T)} \cos (G_1 r). \label{eq:gr_scale}
\end{eqnarray}
The power law decay of correlations in real space, oscillating on scales of the dislocation spacing $D$, are similar to 2d ``quasi-long range order''~\cite{nelson1979dislocation,halperin1978theory}, but arise here in a 1d system with long range interactions. 

\begin{figure}[htb]
    \centering
    \includegraphics[width=0.9\columnwidth]{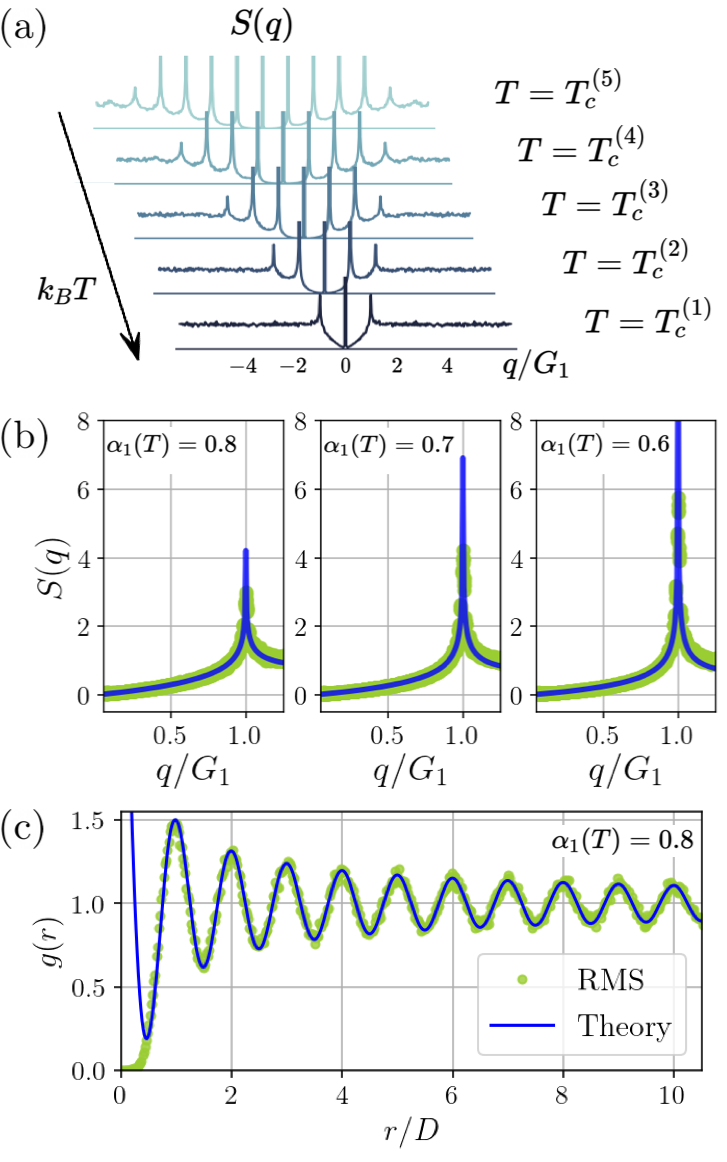}
    \caption{(a) Structure factors extracted from random matrix simulations (RMS) at the critical temperatures in Eq.~(\ref{eq:Tcm}) show the sequential disappearance of the algebraic Bragg peak divergence at $G_m$. (b) The first-order Bragg peak extracted from RMS (green) compared with our theory in Eq.~(\ref{eq:S1}) (blue) at three different temperatures given by inverting $\alpha_1(T)$ in Eq.~(\ref{eq:sq_exp}). Theory and RMS show excellent agreement.}
    \label{fig:LAGB_Sq}
\end{figure}

\textit{Random matrix simulations.}---
We now utilize random matrix theory---the general $\beta$-Gaussian (Hermite) ensemble~\cite{dumitriu2002matrix}---which can efficiently simulate the long range interactions embodied in Eq.~(\ref{eq:H_melt}) at finite temperatures. We can test quantitatively the predictions Eq.~(\ref{eq:Sq_final}) and Eq.~(\ref{eq:gr_scale}) that result from a harmonic approximation to this quantity, because the matrix eigenvalues correspond to dislocation positions. The time scale for direct numerical simulations with, say, molecular dynamics for $N$ particles with long range interactions can be quite large, scaling as $O(N^{5/2})$~\footnote{With long range interactions, all $N$ particles must be updated for each of the $N$ equations of motion, so the computational effort scales as $\sim N^2 \tau $. The time scale $\tau$ to achieve equilibrium is inversely proportional to the frequency of longest wavelength longitudinal phonon modes $\tau^{-1} \sim \omega(q) \sim q^{1/2}  > \sim N^{-1/2}$, giving an equilibriation time that scales like $N^{5/2}$.}. However, we can obtain an equilibrium configuration of a LAGB at any temperature $ T$ by diagonalizing the following symmetric, tridiagonal random matrix~\cite{dumitriu2002matrix}, an operation which scales only as $O(N\log N)$~\cite{coakley2013fast},
\footnotesize   % 
\begin{eqnarray}  \label{eq:Hb}
H_\beta = \frac{1}{\sqrt{2}} \left[ \begin{matrix}
N(0, 2)             & \chi_{(N-1) \beta}&                   &     & 0 \\
\chi_{(N-1) \beta} & N(0, 2)           & \chi_{(N-2) \beta}&     & \\
                    & \ddots            & \ddots            & \ddots  & \\
                    &                   & \chi_{2 \beta}    & N(0, 2) & \chi_{ \beta} \\
0                    &                   &                   & \chi_{ \beta}  & N(0, 2) \end{matrix} \right]. 
\end{eqnarray}\normalsize   % 
In Eq.~(\ref{eq:Hb}), $N(0,2)$ indicates a random number drawn from the normal distribution with mean 0 and variance 2, $\chi_{k}$ represents a random number drawn from the chi distribution~\cite{johnson1995continuous}, and $\beta>0$ can assume \textit{any} positive value. 
We can easily tune the temperature of our simulations by changing the random matrix inverse temperature parameter $\beta$, i.e. the Dyson index, which is related to the inverse temperature of the LAGB by
\begin{equation} \label{eq:betas}
\beta \equiv\frac{Yb^2}{4 \pi} \frac{1}{k_B T}. 
\end{equation}
The eigenvalue density of Eq.~(\ref{eq:Hb}) follows the Wigner semicircular distribution~\cite{dumitriu2002matrix}, while the eigenvalue statistics near the center of the spectrum (with an approximately flat density of states) map exactly onto the statistical mechanics of  LAGBs (see Supplemental Material for details~\cite{Note1}). Specifically, the joint probability distribution function (JPDF) of the random matrix eigenvalues at the center of the spectrum at a particular value of $\beta$ is proportional to the dislocation Boltzmann factor $e^{-\frac{H}{k_B T}}$ associated with Eq.~(\ref{eq:H_melt}) for a LAGB at temperature $T$ corresponding to Eq.~(\ref{eq:betas})~\footnote{Note that we do \textit{not} set $\beta = 1/k_B T$, the usual notational convention in statistical mechanics.}. 

The structure factors $S(q)$ extracted from random matrix simulations in Fig.~\ref{fig:LAGB_Sq}a indeed reveal the sequential disappearance of the power law divergence of the $m$-th Bragg peaks at the transition temperatures $\{T_c^{(m)}\}$ predicted by Eq.~(\ref{eq:Tcm}). 
We also checked that the Bragg peak divergence in $S(q)$ and the decay of the radial distribution function $g(r)$ behave according to Eqs.~(\ref{eq:Sq_0}), (\ref{eq:Sq_final}), and (\ref{eq:gr_scale}), with temperature-dependent exponents given by $1-\alpha_m(T)$ and $\alpha_m(T)$, respectively. 

We can also utilize results from random matrix theory for $\textit{standard}$ Gaussian ensembles~\cite{mehta2004random} to conjecture exact asymptotic expressions for $S(q)$ near the first Bragg peak in the temperature range $T_c^{(2)} < T \leq T_c^{(1)} $. Upon combining the scalings embodied in Eqs.~(\ref{eq:Sq_0}) and (\ref{eq:Sq_final}), we expect that (see Supplemental Material~\cite{Note1}),
\footnotesize  
\begin{eqnarray} 
S(q) &\approx &  \frac{\alpha_1(T)}{2} |\bar q| +\left| \frac{\bar q}{2} \right|^{\alpha_1(T)} \frac{\alpha_1(T) }{2(1-\alpha_1(T))} \left[ \frac{1}{ ( 1-\bar q)^{1 - \alpha_1(T)}} - 1 \right],\nonumber\\ \label{eq:S1}
\end{eqnarray}
\normalsize  
where $\bar q = \frac{q}{G_1}$, with analogous results for the radial distribution function.
% from which we determine the coefficients of Eq.~(\ref{eq:gr_scale}),
% \begin{eqnarray} \label{eq:gr_fit}
% \lim_{r \rightarrow \infty} g(r) = 1 + 2 \frac{\cos\left (2 \pi r/D\right )}{\left( 8 r/D\right)^{\alpha_1(T)}}. 
% \end{eqnarray}
% One can verify that Eq.~(\ref{eq:S1}) indeed diverges as the appropriate power law in Eq.~(\ref{eq:Sq_final}) near the first Bragg peak $\bar q \rightarrow 1$, and is consistent with our theory in the $q \rightarrow 0$ limit in Eq.~(\ref{eq:Sq_0}). 
As shown in Fig.~\ref{fig:LAGB_Sq}b, Eq.~(\ref{eq:S1}) shows excellent agreement with results from random matrix simulations.  

%%%%%%%%%%%%Conclusion%%%%%%%%%%%%%%%%%%%%%%
We note in conclusion that in the absence of a coexisting 3d vapor, dislocation climb out of the glide plane can be frozen out at temperatures far below the melting transition of the host crystal $T\ll T_m$~\cite{moretti2004depinning}. The thermal depinning transition associated with the transverse dislocation glide modes nevertheless occurs as described above. Above the depinning transition, as temperature increases, dislocation climb will be increasingly facilitated by dislocations created and annihilated via glide motion to and from the boundaries of the host crystal and by the proliferation of dislocation pairs near the melting of the host lattice~\cite{chui1983grain,fisher1979defects}. Although the LAGB structure factor might approximate delta function Bragg peaks when $T \ll T_m$ as climb is forbidden, we expect a gradual crossover to algebraic Bragg peaks at higher temperatures such that $T \gtrsim T_P$. Phase field methods~\cite{elder2007phase,mellenthin2008phase} might be a particularly efficient way of testing our results via simulations. 

In the future, we hope to obtain a similar understanding of the statistical mechanics of both grain boundaries and dislocation pileups subject to a Peierls potential in three dimensions~\cite{moretti2004depinning,liao2018grain} and two-dimensional materials that allow out-of-plane deformations~\cite{yazyev2010topological}.

\begin{acknowledgements}
We are grateful for helpful conversations with F. Spaepen and J. Huang. G.H.Z. acknowledges support by the Paul and Daisy Soros Fellowship and the National Science Foundation Graduate Research Fellowship under Grant No. DGE1745303. This work was also supported by the NSF through the Harvard Materials Science and Engineering Center, via Grant No. DMR-2011754, as well as by Grant No. DMR-1608501.
\end{acknowledgements}

\bibliography{references.bib}
\clearpage


\section*{Supplementary material}

# Supplemental Material: Statistical Mechanics of Low Angle Grain Boundaries in Two Dimensions

Grace H. Zhang and David R. Nelson
*Department of Physics, Harvard University, Cambridge, MA 02138, USA.*

(Dated: November 9, 2020)

Here, we detail the calculations showing that grain boundaries are ground state features of flat two-dimensional (2d) crystals with trapezoidal boundary conditions (Sec. I), the random matrix mapping between low angle grain boundaries (LAGBs) and the general $\beta$-Gaussian ensemble (Sec. II), and the derivations for the asymptotic expressions of the structure factor and pair correlation function above the thermal depinning transition (Sec. III).

## I. GRAIN BOUNDARIES AS EQUILIBRIUM FEATURES OF FLAT 2D CRYSTALS

In this section, we briefly show that flat 2d crystals with alternative boundary conditions, such as the trapezoidal boundaries shown in Fig. 1a in the main text, can exhibit at least one grain boundary in their equilibrium ground states. The situation here is somewhat similar to an Abrikosov flux lattice in a 2d superconductor [1], in that the strain imposed by the boundary conditions acts like an external magnetic field. Here, however, the anisotropic defect interactions lead to a linear structure as opposed to a 2d defect lattice.

Let us consider a crystal with trapezoidal boundary conditions, as shown in Fig. 1a of the main text. Such a wedge without any defects would experience an appreciable nonzero strain field throughout the material proportional to the grain boundary angle squared. In fact, the strain will be $u_{ij} \sim \theta$, where $\theta$ is the misorientation angle of the grain boundary. The energy of such a defect-free wedge scales as

$$E_1(\theta) \sim \frac{1}{2} Y \theta^2 L W \tag{1}$$

where $L$ and $W$ are the length and width of the crystal (along the horizontal and vertical dimension in Fig. 1a, respectively), and $Y$ is the 2d Young's modulus. We assume $L \sim W$.

However, for a wedge bisected by a grain boundary, composed of dislocations with spacing D, the strain from each dislocation rises to only $u_{ij} \sim b/D \sim \theta$ at distance $D$ from the boundary. Here, the dislocation Burgers vector $b$ equals the host crystal lattice constant $a$. Beyond the distance $D$ from each dislocation, the strain becomes screened by the other dislocations: the anisotropic strain fields of the dislocations cancel out along the vertical direction for distances larger than $D$. The energy contributed by each dislocation is thus of order $\frac{1}{8\pi} Y b^2 \ln(D/a)$ and the total energy of this configuration is thus

$$E_2(\theta) \sim N_d \frac{1}{8\pi} Y b^2 \ln\left(\frac{D}{a}\right) = \frac{YLb}{8\pi} \theta \ln\left(\frac{1}{\theta}\right) \tag{2}$$

where $N_d = L/D$ is the total number of dislocations in the grain boundary.

Hence, when $E_2(\theta) < E_1(\theta)$, which corresponds to

$$\frac{\theta}{\ln(1/\theta)} > \frac{b}{4\pi W} \tag{3}$$

or roughly,

$$\theta \gtrsim \frac{b}{4\pi W} \ln\left(\frac{4\pi W}{b}\right), \tag{4}$$

the wedge with a grain boundary has a lower energy than a defect-free wedge. This is a very easy criterion to satisfy in both simulations and experiments, even with small $\theta \ll 1$. Thus, for a sufficiently large crystal with trapezoidal boundary conditions and $W \gg a$, the grain boundary indeed exists in the equilibrium ground state configuration of the crystal.

## II.  MAPPING BETWEEN LAGBS AND THE GENERAL $\beta$-GAUSSIAN ENSEMBLE

Matrix models of the *general* $\beta$-Gaussian ensembles, where $\beta$ assumes any positive real value, take the following real symmetric *tridiagonal* form [2] shown in Eq. (19) in the main text,

$$H_\beta = \frac{1}{\sqrt{2}} \begin{bmatrix} N(0,2) & \chi_{(N-1)\beta} & & & 0 \\ \chi_{(N-1)\beta} & N(0,2) & \chi_{(N-2)\beta} & & \\ & \ddots & \ddots & \ddots & \\ & & \chi_{2\beta} & N(0,2) & \chi_\beta \\ 0 & & & \chi_\beta & N(0,2) \end{bmatrix}, \tag{5}$$

where $N(0,2)$ indicates a random number drawn from the normal probability distribution with mean 0 and variance 2, $\chi_k$ indicates a random number drawn from the chi distribution, which describes the statistics of $\sqrt{\sum_{i=1}^k Z_i}$, where $Z_1, \cdots, Z_k$ are $k$ independent normally distributed variables with mean 0 and variance 1. The probability density function $p_k(x)$ corresponding to the chi distribution $\chi_k$ is then [3]

$$p_k(x) = \begin{cases} \frac{x^{k-1} e^{-x^2/2}}{2^{k/2-1} \Gamma(\frac{k}{2})}, & x \geq 0 \\ 0, & \text{otherwise,} \end{cases} \tag{6}$$

where $\Gamma(\frac{k}{2})$ is the gamma function, and $k$ does not have to be an integer and can in fact assume any real value. Crucially, unlike for the classical Gaussian random matrices, the Dyson index $\beta > 0$ can now assume *any* positive value. Note that all diagonal elements $H_{\beta,ii}$ are *independently* drawn from $N(0,2)$, whereas each $H_{\beta,ij} = H_{\beta,ji}$ $(i \neq j)$ off-diagonal pair are in fact the same number, so these matrices are symmetric with real eigenvalues.

Upon rescaling the $N$ eigenvalues $(x_1, \cdots, x_N)$ according to $x_i \to \sqrt{2\beta N} x_i$, so that the spectrum lies in the interval $x \in (-1, 1)$, the eigenvalue JPDF is, up to a normalization constant [2],

$$n(x_1, \cdots, x_N) \sim e^{-\beta N \left[ \sum_i x_i^2 - \frac{1}{2N} \sum_{j \neq k} \ln |x_j - x_k| \right]} \equiv e^{-\bar{H}}. \tag{7}$$

In the large $N$ limit, the average eigenvalue density distribution $n(x) = \int dx_2 \cdots dx_N \, n(x_1, \cdots, x_N)$ is given by Wigner's semicircle law [2, 4],

$$n(x) = \frac{2N}{\pi} \sqrt{1 - x^2}. \tag{8}$$

Near the center of the spectrum where $x \approx 0$, the average eigenvalue density becomes approximately uniform

$$n(x \approx 0) \approx \frac{2N}{\pi}, \tag{9}$$

and the one-body potential in $\bar{H}$, which is of order $O(x_i^2)$, can be neglected. Upon substituting in $\beta = Yb^2/4\pi k_B T$ (see also Eq. (20) in main text), the exponential term in Eq. (7) that weights the eigenvalues near the center of the spectrum becomes

$$\bar{H}(\{x_i \approx 0\}) \approx -\frac{1}{k_B T} \frac{Yb^2}{8\pi} \sum_{j \neq k} \ln |x_j - x_k|, \tag{10}$$

which is exactly equal to the reduced Hamiltonian $H/k_B T$ of LAGBs, with transverse fluctuations traced out, displayed in Eq. (9) of the main text.

## III. ASYMPTOTIC FORMS FOR $S(q)$ AND $g(r)$

In contrast to the general $\beta$ matrices, the usual classical $\beta$-Gaussian matrices [5], with eigenvalue statistics identical to their general $\beta$ counterparts, only allow $\beta$ to assume three possible values $\beta = 1, 2, 4$ [6]. Nevertheless, the analytical results derived via orthogonal polynomials for these standard random matrix ensembles ($\beta = 1, 2, 4$) [5] are useful for deriving an exact asymptotic form of the structure factor as a function of the longitudinal momenta $q_x$ [7].

As discussed in the main text, the dimensionless inverse temperature parameter $\beta$ in the Gaussian random matrix ensemble is given by $\beta = Yb^2/4\pi k_B T$ (see also Eq. (20) in main text). The temperature dependent quantity $\alpha_m(T) = m^2 \frac{16\pi k_B T}{Yb^2}$ (Eq. (15) in main text) can then be written in terms of this random matrix parameter $\beta$ as

$$\alpha_m(T) = \frac{4m^2}{\beta}. \tag{11}$$

### A. Structure factor $S(q)$

The structure factor in the $q \to 0$ and $q \to G_m$ limits from the main text can also be expressed in terms of $\beta$ as

$$S(q) \stackrel{q \to 0}{=} \frac{2}{\beta} \frac{q}{G_1} \tag{12}$$

$$S(q) \stackrel{q \to G_m}{\sim} \frac{1}{|G_m - q|^{1 - \frac{4m^2}{\beta}}}. \tag{13}$$

Note that the $m$-th order Bragg peak, centered at $G_m = m\frac{2\pi}{D}$ for a uniform lattice, disappears when $\beta < \beta_c^{(m)}$, with

$$\beta_c^{(m)} = 4m^2. \tag{14}$$

Conveniently, the exact form of the structure factor near the band center at three specific temperatures $\beta = 1, 2, 4$ for the uniform lattice (derived by studying the semicircular Wigner distribution of eigenvalues in the limit of infinite length) can be obtained from random matrix theory using orthogonal polynomials [5]. These results are summarized in Table I, where $\bar{q} \equiv q/G_1$ such that the $m$-th Bragg peak is centered at $\bar{q} = m$. We can see from Table I that the random matrix theory results at $\beta = (1, 2, 4)$ ($T = (\frac{1}{4}, \frac{1}{2}, 1) \times T_c^{(1)}$) are consistent with our theory in the $q \to 0$ limit (Eq. (12)):

$$\lim_{q \to 0} S(q) \approx \frac{2}{\beta} |\bar{q}|. \tag{15}$$

In particular, the results from random matrix theory at $\beta = 4$ motivate us to propose an exact asymptotic expression for the structure factor near the first Bragg peak. In the temperature range $T_c^{(1)} > T > T_c^{(2)}$, only the Bragg peak divergence at the first reciprocal lattice vector $G_1$ remains, and we can approximate the structure factor by

$$S(q) \approx S_0(q) + S_1(q), \tag{16}$$

where $S_0(q)$ dominates near $q \to 0$ in Eq. 15, and $S_1(q)$ dominates in the limit $q \to G_1$, quantities we shall now determine.

To determine the exact form for $S_1(q)$ in Eq. (16), we require that the following conditions are satisfied: (1) $S_1(q)$ is consistent with Eq. (13) in the $q \to G_1$ limit, (2) $S_1(q)$ is subdominant to $S_0(q)$ in the $q \to 0$ limit, and (3) $S(q) = S_0(q) + S_1(q)$ reduces to the exact result from random matrix theory at $\beta = 4$ ($\alpha_1(T) = 1$). Based on these three conditions, we conjecture that the contribution due to the first Bragg peak $S_1(q)$ can be written as

$$S_1(q) = \left|\frac{\bar{q}}{2}\right|^{\alpha_1(T)} \frac{\alpha_1(T)}{2(1 - \alpha_1(T))} \left[\frac{1}{(1 - \bar{q})^{1 - \alpha_1(T)}} - 1\right]. \tag{17}$$

One can verify that Eq. (17) indeed satisfies the three conditions listed above. First, Eq. (17) indeed diverges as the appropriate power law near the first Bragg peak $\bar{q} \to 1$. This is apparent for $\alpha_1 > 1$. In the limit of $\alpha_1 \to 1$, one can use the following identity

$$\lim_{p \to 0} \frac{1}{p} \left(\frac{1}{|k|^p} - 1\right) = -\ln|k| \tag{18}$$

to see that $S_1(q)$ diverges logarithmically as $q \to G_1$. Second, in the limit of $q \to 0$, $S_1(q)$ scales as

$$\lim_{\substack{\alpha_1 \to 1, \\ q \to 0}} S_1(q) = O(|\bar{q}|^{1+\alpha_1(T)}), \quad (19)$$

which is subdominant to $S_0(q) \sim k_B T |\bar{q}|$ provided $T > 0$ (recall that $\bar{q} = q/(\frac{2\pi}{D})$). Finally, for $\alpha_1(T) = 1$ ($T = T_c^{(1)}$ and $\beta = 4$), Eq. (17) reduces to the following

$$\lim_{\beta \to 4} S_1(q) = \frac{|\bar{q}|}{4} \ln |1 - |\bar{q}||, \quad (20)$$

which matches the exact result from random matrix theory in row 3 of Table I.

Upon combining Eqs. (12), (17) and (11), our conjectured form for the structure $S(q)$ in the temperature range $T_c^{(2)} < T \leq T_c^{(1)}$ ($4 \leq \beta < 16$) can be expressed in terms of $\beta$ as

$$S(q) = \frac{2}{\beta} \left[ |\bar{q}| + \left|\frac{\bar{q}}{2}\right|^{4/\beta} \frac{1}{1 - (4/\beta)} \left( \frac{1}{(1-\bar{q})^{1-(4/\beta)}} - 1 \right) \right], \quad (21)$$

which is the result Eq. (21) in the main text.

| $\beta$ | $S(q) \equiv K(\bar{q}), \quad \bar{q} \equiv \frac{q}{G_1}$ |
|---|---|
| 1 | $2|\bar{q}| - |\bar{q}| \ln(1 + 2|\bar{q}|), \quad |\bar{q}| \leq 1$ |
|   | $2 - |\bar{q}| \ln \left( \frac{2|\bar{q}|+1}{2|\bar{q}|-1} \right), \quad |\bar{q}| \geq 1$ |
| 2 | $|\bar{q}|, \quad |\bar{q}| \leq 1$ |
|   | $1, \quad |\bar{q}| \geq 1$ |
| 4 | $\frac{1}{2}|\bar{q}| - \frac{1}{4}|\bar{q}| \ln|1 - |\bar{q}||, \quad |\bar{q}| \leq 2$ |
|   | $1, \quad |\bar{q}| \geq 2$ |

TABLE I. Exact expressions for $S(q) \equiv K(\bar{q})$, where $D$ is the dislocation spacing and $\bar{q} = q/G_1 = qD/2\pi$, derived from random matrix theory via orthogonal polynomials for the special values of the dimensionless random matrix inverse temperature parameter $\beta = 1, 2, 4$ [5].

### B. Radial distribution function $g(r)$

The two-point correlation functions from random matrix theory for the uniform lattice (derived by studying the semicircle lattice in the limit of infinite length) at the special dimensionless inverse temperature parameter $\beta = \frac{Y b^2}{4\pi k_B T} = 1, 2, 4$ are shown in Table II, along with their behavior when the separation distance $r$ is large relative to the lattice spacing $D$ [5].

We obtain the shape of the envelope modulating the oscillations in $g(r)$ as

$$\lim_{r \to \infty} (g(r) - 1) \sim \frac{\cos(2\pi r/D)}{(r/D)^{\alpha_1(T)}}. \quad (22)$$

Upon utilizing the exact result for $\beta = 4$ ($\alpha_1(T) = 1$) in row 3 of Table II, we determine the coefficients of Eq. (22) up to one fitting parameter $c$:

$$\lim_{r \to \infty} g(r) = 1 + \frac{c \cos(2\pi r/D)}{4 (cr/D)^{\alpha_1(T)}}. \quad (23)$$

Upon examining the entire temperature range $T_c^{(1)} \geq T > T_c^{(2)}$ ($16 > \beta \geq 4$), we find that $c \approx 8.0$, so that the radial

| $\beta$ | $g(r) = h(\bar{r}), \quad \bar{r} = \frac{r}{D}$ |
|---|---|
| 1 | $1 - \left(\int_{\bar{r}}^{\infty} s(\bar{t})d\bar{t}\right)\left(\frac{d}{dr}s(\bar{r})\right) + (s(\bar{r}))^2, \quad s(\bar{r}) = \frac{\sin \pi \bar{r}}{\pi \bar{r}}$ |
| | Large $\bar{r}$: $1 - \frac{1}{\pi^2 \bar{r}^2} + \frac{3}{2\pi^4 \bar{r}^4} + \frac{\cos 2\pi \bar{r}}{\pi^4 \bar{r}^4} + \cdots$ |
| 2 | $1 - s(\bar{r})^2$ |
| | Large $\bar{r}$: $1 - \frac{1}{2\pi^2 \bar{r}^2} + \frac{\cos 2\pi \bar{r}}{2\pi^2 \bar{r}^2} + \cdots$ |
| 4 | $1 - s(2\bar{r})^2 + \frac{d}{d\bar{r}}s(2\bar{r}) \cdot \int_0^{\bar{r}} s(2\bar{t})d\bar{t}$ |
| | Large $\bar{r}$: $1 + \frac{\pi}{2}\frac{\cos(2\pi\bar{r})}{2\pi\bar{r}} + \cdots$ |

TABLE II. Exact expressions for $g(r) = h(\bar{r})$ from random matrix theory derived via orthogonal polynomials, where $\bar{r} = \frac{r}{D}$ scales the inter-eigenvalue distance $r = |x_1 - x_2|$ by the mean eigenvalue spacing near the center of the semicircle lattice $D$, and $s(\bar{r}) \equiv \frac{\sin \pi \bar{r}}{\pi \bar{r}}$ [5].

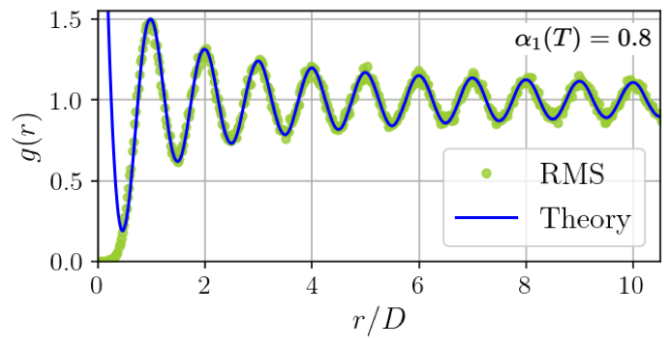

FIG. 1. Radial distribution function $g(r)$ extracted from random matrix simulations (RMS) (green) and Eq. (24) (blue) at temperature $T$ corresponding to $\alpha_1(T) = 0.8$. Theory and simulations show good agreement.

distribution function at large $r$ takes the following form

$$\lim_{r \to \infty} g(r) = 1 + 2\frac{\cos(2\pi r/D)}{(8r/D)^{\alpha_1(T)}}. \tag{24}$$

Fig. 1 shows good agreement between Eq. (24) and random matrix simulations.


\end{document}